\pdfoutput=1
\RequirePackage[l2tabu,orthodox]{nag}
\documentclass[a4paper,11pt]{article}
\usepackage[T1]{fontenc}
\usepackage[utf8]{inputenc}
\usepackage{pos}
\usepackage{microtype}
\usepackage[artemisia]{textgreek}
\usepackage{mathtools}
\usepackage{csquotes}
\usepackage{physics}
\usepackage{booktabs}
\usepackage[nolist]{acronym}
\usepackage{siunitx}
\usepackage{tikz}
\usepackage{pgf}

\usepackage[style=base]{caption}
\usepackage{subcaption}
\NewDocumentCommand\cpi{}{\text{\textpi}}                        

\NewDocumentCommand\aHVP{}{a_\mu^{\mathrm{HVP,LO}}}

\NewDocumentCommand\Dalphahad{}{\Delta\alpha_{\mathrm{had}}}

\NewDocumentCommand\dalpfive{ m }{\Delta\alpha^{(5)}_{\mathrm{had}}(#1)}
\NewDocumentCommand\SPi{}{\bar{\Pi}}

\DeclareSIUnit\fm{\femto\metre}

\begin{acronym}
  \acro{SM}{Standard Model}
  \acro{BSM}{beyond the Standard Model}
  \acro{QFT}{quantum field theory}
  \acro{QED}{quantum electrodynamics}
  \acro{QCD}{quantum chromodynamics}
  \acro{pQCD}{perturbative QCD}
  \acro{RG}{renormalization group}
  \acro{MC}{Monte Carlo}
  \acro{MDU}{molecular dynamics unit}
  \acro{HVP}{hadronic vacuum polarization}
  \acro{HLbL}{hadronic light-by-light scattering}
  \acro{TMR}{time-momentum representation}
  \acro{CCS}{Lorentz-covariant coordinate space}
  \acro{FFT}{fast Fourier transform}
  \acro{CLS}{Coordinated Lattice Simulations}
  \acro{BC}{boundary condition}
  \acro{OBC}{open boundary condition}
  \acro{PBC}{periodic boundary condition}
  \acro{GS}{Gounaris-Sakurai}
  \acro{SN}[S/N]{signal-to-noise ratio}
  \acro{chiPT}[\textchi PT]{chiral perturbation theory}
  \acro{NLO}{next-to-leading order}
  \acro{NNLO}[N\textsuperscript{2}LO]{next-to-next-to-leading order}
  \acro{HP}{Hansen-Patella}
	\acro{FSE}{finite-size effects}
	\acro{MLLGS}[MLL-GS]{Meyer-Lellouch-Lüscher Gounaris-Sakurai}
	\acro{SM}{Standard Model}
	\acro{BMWc}{Budapest-Marseille-Wuppertal collaboration}
\end{acronym}

\title{The hadronic running of the electroweak couplings from lattice QCD}

\author*[a]{Marco~Cè}
\affiliation[a]{Albert Einstein Center for Fundamental Physics (AEC) and Institut für Theoretische Physik, Universität Bern, Sidlerstrasse 5, 3012 Bern, Switzerland}
\emailAdd{marcoce@itp.unibe.ch}

\author[b]{Antoine~Gérardin}
\author[c]{Georg~von~Hippel}
\author[c,d]{Harvey~B.~Meyer}
\author[c,d,e]{Kohtaroh~Miura}
\author[c]{Konstantin~Ottnad}
\author[f]{Andreas~Risch}
\author[g]{Teseo~San~José}
\author[c,d]{Hartmut~Wittig}
\affiliation[b]{Aix-Marseille Université, Université de Toulon, CNRS, CPT, Marseille, France}
\affiliation[c]{PRISMA\textsuperscript{+} Cluster of Excellence and Institut für Kernphysik, Johannes Gutenberg-Universität Mainz, 55099 Mainz, Germany}
\affiliation[d]{Helmholtz-Institut Mainz, Johannes Gutenberg-Universität Mainz, 55099 Mainz, Germany\\
                and GSI Helmholtzzentrum für Schwerionenforschung, Planckstraße 1, 64291 Darmstadt, Germany}
\affiliation[e]{KEK Theory Center, High Energy Accelerator Research Organization, Tsukuba, Japan}
\affiliation[f]{John von Neumann-Institut für Computing NIC, Deutsches Elektronen-Synchrotron DESY, Platanenallee 6, 15738 Zeuthen, Germany}

\affiliation[g]{Laboratoire de Physique des 2 Infinis Irène Joliot-Curie, CNRS/IN2P3, Université Paris-Saclay, 91405 Orsay, France}

\abstract{
The energy dependency (running) of the strength of electromagnetic interactions $\alpha$ plays an important role in precision tests of the Standard Model. The running of the former to the $Z$ pole is an input quantity for global electroweak fits, while the running of the mixing angle is susceptible to the effects of Beyond Standard Model physics, particularly at low energies.
We present a computation of the hadronic vacuum polarization (HVP) contribution to the running of these electroweak couplings at the non-perturbative level in lattice QCD, in the space-like regime up to $Q^2$ momentum transfers of $7\,\mathrm{GeV}^2$. This quantity is also closely related to the HVP contribution to the muon $g-2$.
We observe a tension of up to \num{3.5} standard deviation between our lattice results for $\Delta\alpha^{(5)}_{\mathrm{had}}(-Q^2)$ and estimates based on the $R$-ratio for $Q^2$ in the $3$ to $7\,\mathrm{GeV}^2$ range. The tension is, however, strongly diminished when translating our result to the $Z$ pole, by employing the Euclidean split technique and perturbative QCD, which yields $\Delta\alpha^{(5)}_{\mathrm{had}}(M_Z^2)=0.027\,73(15)$. This value agrees with results based on the $R$-ratio within the quoted uncertainties, and can be used as an alternative to the latter in global electroweak fits.
}

\FullConference{%
  41st International Conference on High Energy physics - ICHEP2022\\
  6-13 July, 2022\\
  Bologna, Italy
}

\begin{document}
\maketitle

\section{Introduction}

In the \ac{SM} of particle physics, the strength of electromagnetic interactions is represented by the \ac{QED} coupling $\alpha$ that depends on the energy at which electromagnetism is probed.
At low energies the value of $\alpha^{-1}=\num{137.035 999 139(31)}$  is known experimentally to better than a part per billion.
At higher energies, such as at the $Z$ pole, the most precise determination is obtained by applying the renormalization group running.
Choosing to work in the \emph{on-shell} scheme, one has
\begin{equation}
  \alpha(q^2) = \frac{\alpha}{1-\Delta\alpha(q^2)} ,
\end{equation}
where the major contribution to the error on the running $\Delta\alpha(q^2)$ comes from the low-energy hadronic component $\Dalphahad(q^2)$, which is proportional to the subtracted 
\ac{HVP} function of an off-shell photon propagator
\begin{equation}
  \Dalphahad(q^2) = 4\cpi\alpha \Re\SPi(q^2) \qquad \SPi(q^2) = \Pi(q^2) - \Pi(0) .
\end{equation}
A precise determination of the \ac{HVP} function and in turn of $\Dalphahad(q^2)$ can be obtained from experimental hadronic cross-section data encoded in the well-known ratio $R(s)$~\cite{Keshavarzi:2018mgv,Davier:2019can,Jegerlehner:2019lxt}.
Alternatively, the HVP function $\SPi(-Q^2)$ at space-like $Q^2>0$ momentum transfers can be computed on the lattice~\cite{Burger:2015lqa,Francis:2015grz,Borsanyi:2017zdw,Ce:2022eix}.
In combination with the Euclidean split technique~\cite{Jegerlehner:2008rs}, this can be used to provide a value of $\Dalphahad(M_Z^2)$ independent of $R(s)$ data, which is the main result presented here~\cite{Ce:2022eix}.

The interests in computing $\SPi(-Q^2)$ on the lattice is further motivated by the connection with the anomalous magnetic moment ($g-2$) of the muon, and the long-standing tension between the muon $g-2$ theory prediction and experimental results~\cite{Aoyama:2020ynm}.
In particular, the leading-order muon $g-2$ \ac{HVP} contribution can be expressed as an integral over $\SPi(-Q^2)$ at space-like $Q^2$ with a positive-definite integration kernel
\begin{equation}
\label{eq:aHVP_integral}
  \aHVP = 4\alpha^2 \int_0^1 \dd{x}(1-x) \SPi(-Q^2) , \qquad Q^2 = \frac{x^2m_\mu^2}{1-x} ,
\end{equation}
a fact that is at the core of the MUonE experiment proposal for an independent measurement of $\aHVP$ in $t$-channel scattering experiments~\cite{Abbiendi:2016xup}.
The precise lattice determination of $\aHVP$ by the \ac{BMWc}~\cite{Borsanyi:2020mff} points to a larger value---in tension with the data-driven determination and closer to solving the $g-2$ puzzle---that according to eq.~\eqref{eq:aHVP_integral} has to be correlated with an increase of $\SPi(-Q^2)$ and $\Dalphahad(-Q^2)$.
The evidence is even stronger when lattice data for the so-called \ac{HVP} window is considered~\cite{Ce:2022kxy,Alexandrou:2022amy}.
At the same time, this increase implies an increase of $R(s)$ that puts these lattice results in tension with the experimental measurements of the hadronic cross-section.

Computing $\SPi(-Q^2)$ from first principles on the lattice can help shed light on this tension, albeit with limited energy resolution on $R(s)$ due to the lattice data being limited to space-like $Q^2$.
Moreover, the lattice determination of $\Dalphahad(M_Z^2)$ can be used as an alternative input to the data-driven estimate in global EW fits.

\section{Lattice calculation}

The lattice computation is performed using the \ac{TMR} method~\cite{Bernecker:2011gh}.
For each selected value of $Q^2$ up to \SI{7}{\GeV\squared}, we compute the integral
\begin{equation}
  \SPi(-Q^2) = \int_0^\infty \dd{t} G(t) \left[ t^2 - \frac{4}{Q^2} \sin[2](\frac{Qt}{2}) \right] ,
\end{equation}
where the continuum expression for the kernel is used and $G(t)$ is the Euclidean-space correlator of two electromagnetic currents at imaginary time separation $t$ computed on the lattice
\begin{equation}
  G(t) = -\frac{1}{3} \int\dd[3]{x} \sum_{k=1}^3 \ev{j_k(t, \vec{x}) j_k(0)} , \qquad j_\mu = \frac{2}{3}\bar{u}\gamma_\mu u - \frac{1}{3}\bar{d}\gamma_\mu d - \frac{1}{3}\bar{s}\gamma_\mu s + \frac{2}{3}\bar{c}\gamma_\mu c + \dots .
\end{equation}

We use a set of $N_{\mathrm{f}}=2+1$ dynamical quark ensembles of gauge field configurations from the \ac{CLS} effort~\cite{Bruno:2014jqa}.
On these ensembles, the up and down quarks are mass degenerate and the charm quark contribution is present in the valence sector only.
The lattice current is renormalized and both the action and the observables are non-perturbatively $\order{a}$-improved to ensure discretization effects starting at $\order*{a^2}$, where $a$ is the lattice spacing.
We refer the reader to ref.~\cite{Ce:2022eix} for more details.

Our set of ensembles covers four lattice spacings from $a\approx\SI{0.086}{\fm}$ to $\approx\SI{0.050}{\fm}$ and a range of pion and kaon masses from the $\mathrm{SU}(3)$-flavour-symmetric point at $m_\pi=m_K\approx\SI{415}{\MeV}$ to physical ones along a trajectory that keeps $m_K^2+m_\pi^2/2$ approximately constant.
This allow us to reliably extrapolate $\Dalphahad(-Q^2)=4\cpi\alpha\SPi(-Q^2)$ for a set of values of $Q^2$ to the ``physical'' point corresponding to an isospin-symmetric world.

At $Q^2=\SI{5}{\GeV\squared}$, the result at the physical point is
\begin{equation}
\label{eq:lattice_result_5GeV2}
  \Dalphahad(-\SI{5}{\GeV\squared}) = \num{0.00716(9)} .
\end{equation}
On top of the statistical error on the lattice data, we include in the final error budget a systematic uncertainty.
This is composed of the systematic error from varying the fit model, the scale-setting error, the systematics from missing charm sea-quark loops, and the one from missing isospin-breaking effects.

\section{Comparison with phenomenology}
\enlargethispage*{\baselineskip}

\begin{figure}[t]
  \centering
  \scalebox{.5}{\input{figures/running_comparison_ratio.pgf}\input{figures/running_comparison_detail.pgf}}
  \caption{%
    Left, upper panel: ratio of the hadronic running $\Dalphahad$ computed by \ac{BMWc}~\cite{Borsanyi:2017zdw} divided by our results, for five different momenta.
    Left, lower panel: the total hadronic running $\Dalphahad^{(5)}$ from various phenomenological estimates~\cite{Davier:2019can,Jegerlehner:2019lxt,Keshavarzi:2018mgv} and the lattice result of ref.~\cite{Borsanyi:2017zdw}, normalized by the result of this work.
    Right: Compilation of results for the four-flavor $\Dalphahad$ lattice computations~\cite{Borsanyi:2017zdw,Borsanyi:2020mff} (above) and the five-flavor $\Dalphahad^{(5)}$ phenomenological estimates (below) at selected values of $Q^2$.
  }\label{fig:running_comparison_alpha}
\end{figure}

In figure~\ref{fig:running_comparison_alpha} our lattice results are compared to other lattice results from \ac{BMWc} (upper part)~\cite{Borsanyi:2017zdw,Borsanyi:2020mff}, as well as with three data-driven estimates (lower part)~\cite{Davier:2019can,Jegerlehner:2019lxt,Keshavarzi:2018mgv}.
We observe a small tension with the \ac{BMWc} results, especially in the $I=1$ component at small value of $Q^2$, while at larger $Q^2$ the results are mostly compatible.
The tension is instead significant when our result is compared with the data-driven estimates, up to \num{3.5} standard deviations.


\begin{figure}[t]
  \centering
  \scalebox{.5}{\input{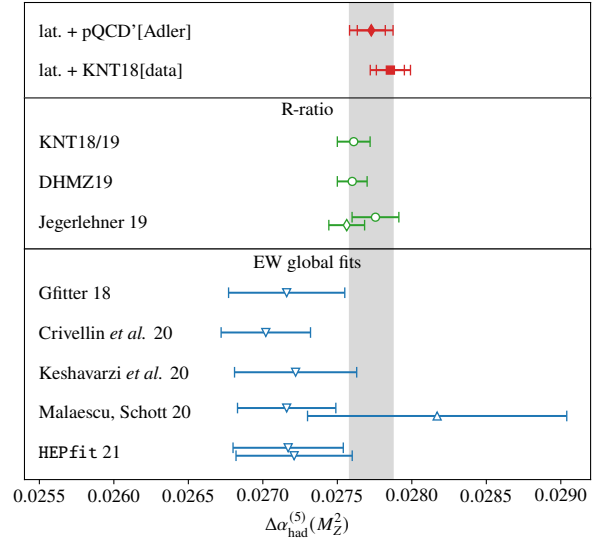}}
  \caption{%
    Compilation of results for $\dalpfive{M_Z^2}$.
    The first two data points (red symbols) are the lattice results of ref.~\cite{Ce:2022eix}.
    Green circles denote results based the data-driven method in, from top to bottom, refs.~\cite{Keshavarzi:2018mgv,Keshavarzi:2019abf}, \cite{Davier:2019can}, and \cite{Jegerlehner:2019lxt}.
    The estimate based on the Adler function in ref.~\cite{Jegerlehner:2019lxt} is shown as a green diamond.
    Blue symbols represent the results from global EW fits, published in refs.~\cite{Haller:2018nnx,Crivellin:2020zul,Keshavarzi:2020bfy,Malaescu:2020zuc,deBlas:2021wap}.
    The upper triangle point from ref.~\cite{Malaescu:2020zuc} does not use the Higgs mass.
    The gray band represents our final result quoted in eq.~\eqref{eq:alp5_mz_lat_padler}.
  }\label{fig:alp5}
\end{figure}

To estimate $\Dalphahad^{(5)}(M_Z^2)$ we employ the so-called Euclidean split technique (or Adler function approach)~\cite{Jegerlehner:2008rs}, which consists in splitting the contribution to the running to $M_Z$ as
\begin{equation}
\label{eq:euclidean_split}
  \Dalphahad^{(5)}(M_Z^2) = \Dalphahad^{(5)}(-Q_0^2) + [\Dalphahad^{(5)}(-M_Z^2) - \Dalphahad^{(5)}(-Q_0^2)] + [\Dalphahad^{(5)}(M_Z^2) - \Dalphahad^{(5)}(-M_Z^2)] .
\end{equation}
In the r.h.s., the first term is evaluated on the lattice and it is the main result presented in the previous section.
The result at $Q_0^2=\SI{5}{\GeV\squared}$ is given in eq.~\eqref{eq:lattice_result_5GeV2}.
The second term involves the running at space-like momenta up to the $Z$ pole.
This can either be estimated using the data-driven method from the hadronic cross section~\cite{Davier:2019can,Keshavarzi:2018mgv,Jegerlehner:2019lxt}, or for large enough $Q_0^2$ it can be computed in \ac{pQCD}.
The result based on \ac{pQCD} is our preferred choice since it allows us to be independent of experimental $R(s)$ input.
Choosing the threshold energy $Q_0^2=\SI{5}{\GeV\squared}$,
the estimate obtained using the \texttt{pQCDAdler} code\footnote{\url{http://www-com.physik.hu-berlin.de/~fjeger/software.html}} is $[\Dalphahad^{(5)}(-M_Z^2) - \Dalphahad^{(5)}(-Q_0^2)]_{\mathrm{pQCD}} = \num{0.020528(107)}$.
Finally, the third term in the r.h.s.\ of eq.~\eqref{eq:euclidean_split} has been estimated in \ac{pQCD} in ref.~\cite{Jegerlehner:2019lxt} and amounts to a small and precise contribution, $[\Dalphahad^{(5)}(M_Z^2) - \Dalphahad^{(5)}(-M_Z^2)]_{\mathrm{pQCD}}=\num{0.000045(2)}$.

Combining these results, the hadronic contribution to the running of the electromagnetic coupling $\alpha$ at the $Z$ pole is
\begin{equation}
\label{eq:alp5_mz_lat_padler}
  \Dalphahad^{(5)}(M_Z^2)|_{\textrm{lat. + pQCD}} = \num{0.02773}(9)_{\mathrm{lat}}(2)_b(12)_{\mathrm{pQCD}} = \num{0.02773(15)} ,
\end{equation}
where the first error is the uncertainty of our lattice estimate, while the second error accounts for the neglected contribution form bottom quark effects, and the third error is the one on the second term in the r.h.s.\ of eq.~\eqref{eq:euclidean_split}.
The total error is augmented by the maximum deviation obtained by varying the threshold energy $Q_0^2$ in the interval $\SI[separate-uncertainty]{5\pm 2}{\GeV\squared}$, which has a negligible effect.

Our results are compared in figure~\ref{fig:alp5} with other estimates using the standard dispersive approach~\cite{Davier:2019can,Keshavarzi:2018mgv,Jegerlehner:2019lxt} and global EW fits~\cite{Haller:2018nnx,Crivellin:2020zul,Keshavarzi:2020bfy,Malaescu:2020zuc,deBlas:2021wap}.
We observe that the tension with the data-driven results is strongly diminished, mostly due to the major contribution to the error of the running from $Q_0^2$ to the $Z$ pole.
Of course, this contribution is fully correlated between lattice and data-driven results, and the tension in the $Q^2\lesssim\SI{5}{\GeV\squared}$ contribution is only hidden.
However, the results from global EW fits show a pull of a bit more than one standard deviation between the indirect determination of $\Dalphahad^{(5)}(M_Z^2)$ and our result in eq.~\eqref{eq:alp5_mz_lat_padler}, indicating no inconsistency between lattice results and the global EW fits, as also observed in ref.~\cite{deBlas:2021wap}.

\section{Conclusions}

We presented a computation of the hadronic contribution to the running of the electromagnetic coupling $\alpha$.
Our result is obtained on the lattice for space-like $Q^2$ up to $\approx\SI{7}{\GeV\squared}$ and it is slightly larger but still compatible with an earlier calculation by \ac{BMWc}.
However, there is a significant tension with the predictions based on the data driven method.

Combining our result obtained in the $Q^2=\SI[separate-uncertainty]{5\pm 2}{\GeV\squared}$ range with \ac{pQCD}, we obtain an estimate for $\Dalphahad^{(5)}(M_Z^2)$ that does not rely on experimental hadronic cross section data as input.
This result is consistent with and of similar precision as estimates employing the data-driven approach.
Moreover, we observe no significant tensions between our lattice result and global EW fits.
\enlargethispage*{3\baselineskip}

\setlength{\bibsep}{0pt}
\bibliography{biblio_bibtex}

\end{document}